\documentclass[prd,preprint]{revtex4}
\usepackage{tabularx}

\usepackage[dvips]{color}
\usepackage[open]{bookmark}
\def\J{\rm J}
\newcommand\E{\rm E}
\def\d{{\rm d}}

\newcommand{\Lg}{\mathcal{L}}

	\usepackage{amsmath}
	\usepackage{amsfonts}
  	\usepackage{amssymb}
  	\usepackage{float}
	\usepackage{makeidx}
	\usepackage{amsfonts}
	\usepackage[ansinew]{}
	\usepackage[usenames,dvipsnames]{pstricks}
	\usepackage{epsfig}

\usepackage{xcolor}

\newcommand{\be}{\begin{equation}}
\newcommand{\ee}{\end{equation}}
\newcommand{\bea}{\begin{eqnarray}}
\newcommand{\eea}{\end{eqnarray}}

\makeindex

\begin{document}

\title{Probing the dark Universe with gravitational waves}
\author{Antonio Enea Romano}

\affiliation{ICRANet, Piazza della Repubblica 10, I--65122 Pescara, Italy}
\affiliation{Instituto de F\'isica, Universidad de Antioquia, A.A.1226, Medell\'in, Colombia}

\date{\today}

\begin{abstract}
Gravitational waves (GW) are expected to interact with  dark energy and dark matter, affecting their propagation on cosmological scales.
In order to model this interaction, we derive a gauge invariant effective equation and action valid for all GWs polarizations, based on encoding the effects of the interaction of GWs at different orders in perturbations, in a polarization, frequency and time dependent  effective speed. The invariance of perturbations under time dependent conformal transformations and  the gauge invariance of the GWs allow to  obtain the unitary gauge effective action in any conformally related frame, making   transparent the relation between Einstein and Jordan frame. 
The propagation time and luminosity distance of different GWs polarizations allow to probe at different frequencies and redshift the dark Universe, which act as an effective medium, whose physical properties can be modeled by the GWs effective speed. 
\end{abstract}
\email{antonioenea.romano@ligo.org}

\pacs{Valid PACS appear here}
\maketitle



\section{Introduction}
The direct observation of GWs by the Laser Interferometer Gravitational Wave Observatory (LIGO) and Virgo has started the era of GW astronomy \cite{LIGOScientific:2016aoc}.
These observations are in good agreement with general relativity (GR) predictions, but even in GR tensors modes are expected to acquire a frequency and polarization dependent effective speed \cite{Romano:2022jeh} due to the interaction with other fields, an effect which can be tested with multimessenger observations in different bands, and with measurements of the luminosity distance of different polarizations modes.

In order to study the effects of the interaction of GWs, we generalize  to all possible GWs polarizations, the effective speed approach \cite{Romano:2022jeh,Romano:2023ozy,Romano:2018frb,Romano:2023uwf}, derived for tensor modes, including gauge invariant scalar and vector modes. We obtain a general effective propagation equation and action which can be applied for model independent analysis of the effects of interaction between different polarizations of GWs, or of GWs with other fields \cite{Romano:2023lxf}, such as dark matter or dark energy. In this effective approach, the other fields, and the other GWs polarizations, act as an effective medium for each GW polarization, which can consequently propagate with a different frequency and time dependent effective speed.
This effective momentum and polarization dependent speed is encoding interaction effects which are conceptually analogous to those encoded in the modified dispersion relation for  electromagnetic waves (EMW) propagating in a plasma.
Given the generality of this effective approach, it is particularly suitable for model independent phenomenological analysis of observational data, and predicts that the speed of gravitational waves $c_{ij}$ and the gravitational luminosity distance can depend on the frequency and polarization of GWs.

\section{Effective speed of arbitrary GW polarization}
 
 GWs $h_{ij}$ propagating in the $z$-direction \cite{Ezquiaga:2018btd}, can be  decomposed  in  different polarizations as  
 \be
 \label{Aij}
 h_{ij}=\begin{pmatrix} h_S+A_{+} & h_\times & h_{V1} \\ h_\times & h_S-h_+ & A_{V2} \\ h_{V1} & h_{V2} & h_L \end{pmatrix}\,,
 \ee
 where $h_+$ and $h_\times$ are the  tensor modes, $h_{V1,2}$ the  vector polarizations, and $h_S$   and $h_L$ the transverse and longitudinal scalar modes. Note that the gauge has not been fully fixed in the above equation, and this can lead to GWs modes which depend on the observer frame \cite{Bonvin:2022mkw}. We will resolve this ambiguity in the next section, in which we will obtain the effective action for  gauge invariant GWs. 

 In order to show the general applicability of the effective speed approach \cite{Romano:2022jeh}, let's consider for example this ansatz for the modified GW propagation equation  \cite{Nishizawa:2017nef}
 \bea
 h''_{ij}+(2+\nu)\mathcal{H} h'_{ij}+(c_g^2k^2+m^2a^2)h_{ij}&=&\Pi_{ij} (h_{ij},h_{pq},\phi_m) \,, \label{GWmod} \\ 
 L_m(\phi_m,h_{ij})&=&0 \,, \label{Lm}
 \eea
 where $\nu$ accounts for a modification of the   friction term, $c_g$  for an modified propagation speed, $m$ is an effective mass, $\Pi_{ij}$ is the source term associated to the self-interaction, interaction with other fields or other GWs polarizations $h_{pq}$,  $\phi_m$ denotes abstractly other fields interacting with GWs, and $L_m$ is the differential operator corresponding to the equation of motion (EOM) of each $\phi_m$.
After rewriting  eq.(\ref{GWmod}) as
 \be
 \label{GWmod2}
 h''_{ij}+2\mathcal{H} h'_{ij}+c^2\, k^2 h_{ij}=\Pi_{ij}-\nu\mathcal{H} h'_{ij}-\Big[(c_g^2-c^2)k^2+m^2a^2\Big]h_{ij}=\Pi_{ij}^{eff}\,,
 \ee
we obtain the effective equation 
\be
h_{ij}''+2 \Big(\frac{a'}{a}-\frac{c'_{ij}}{c_{ij}}\Big) h'_{ij}+k^2 c_{ij}^2 h_{ij}=0 \,, \label{heffij}
\ee
where we have defined \cite{Romano:2022jeh} the effective polarization, momentum and time dependent speed $c_{ij}$ as
\be
c_{ij}^2(\eta,k)=\Big(1-\frac{\hat{g}_{ij}}{\hat{h}'_{ij}}\Big)^{-1} \quad,\quad 
\hat{g}_{ij}=\frac{1}{a^2}\int a^4 \hat{\Pi}^{eff}_{ij}\,d\eta\,, \label{cij}
\ee
In the above equation a hat denotes quantities obtained by substituting the solutions $\hat{h}_{ij},\hat{\phi}_m$ of the system of coupled differential equations given in eqs.(\ref{GWmod}-\ref{Lm}), i.e. $\hat{\Pi}^{eff}_{ij}$ is just a function of space-time coordinates after the substitution, which accounts for the integrated  modified propagation effects. 
Eq.(\ref{cij}) shows that each polarization mode $h_{ij}$ can have a different frequency and time dependent effective speed, depending on the effective anisotropy tensor $\hat{\Pi}^{eff}_{ij}$, evaluated along the propagation path. 

For a set of initial conditions consistent with those used to obtain $\hat{h}_{ij}$, eq.(\ref{heffij}) gives by construction the same solution $\hat{h}_{ij}$. Solutions of eq.(\ref{heffij}) corresponding to  different initial conditions are not physically relevant, such as for example $h_{ij}=0$.
Any solution of eq.(\ref{GWmod}), or any other ansatz which can be manipulated to put it in the form given in eq.(\ref{GWmod2}), can always be obtained as a solution of the effective equation \ref{heffij}, with the effective speed defined in eq.(\ref{cij}). This effective approach is convenient to relate the modified propagation effects to observations, because the GW-EMW luminosity distance ratio is related to the effective speed ratio\cite{Romano:2022jeh}, making the effective equation useful for model independent observational data analysis. 

\section{Einstein frame definition}
In the previous section we have not specified explicitly the frame in which the ansatz (\ref{GWmod}) is given, since our goal was  to give an example of the general applicability  of the effective speed approach. In order to derive a general effective action, without assuming any ad hoc ansatz, it is important to clarify the relation between Jordan and Einstein frame.
We defined the Jordan frame Lagrangian of a modified gravity theory with Jordan frame matter-coupling (JMC) as
\be
\Lg_{\rm JMC}=\sqrt{g_{\J}}\Big[\Omega^2 R_{\J} + L^{\rm MG}_{\J}+L^{\rm matter}_{\J}(g_{\J})\Big]\,\label{JMCJ} \,,
\ee
where $L^{\rm MG}$ and $L^{\rm matter }$ are respectively the modified gravity and matter Lagrangians.
 After performing a conformal transformation $g_{\E}=\Omega^2 g_{\J}$, in the Einstein frame we have 
\be
\Lg_{\rm JMC}=\sqrt{g_{\E}}\Big[R_{\E} + L^{\rm MG}_{\E}+L^{\rm matter}_{\E}(\Omega^{-2} g_{\E})\Big]\label{JMCE}\,.
\ee

Note that while the tensor modes  speed $c_T$ is invariant under conformal transformations,  it is not invariant under disformal transformations, allowing to define a combination of disformal, conformal and coordinate transformations taking to a frame \cite{Creminelli:2014wna} in which $c_T=1$. Note nevertheless that the ratio $c_T/c$ between the speed of gravitational and electromagnetic waves is disformal invariant, implying that, if $c_T/c$ is time dependent,  in the $c_T=1$ frame the speed of light is time dependent. For this reason, while the $c_T=1$ frame is useful to study the resilience of  inflationary predictions\cite{Creminelli:2014wna},  we define the Einstein frame as that in which the coefficient of the Ricci scalar and the  speed of light are constant, and not the $c_T=1$ frame, in agreement with the definition adopted in \cite{Gubitosi:2012hu} for example. 

\section{Gauge invariant gravitational waves}
For the purpose of understanding the role played by gauge transformation it is convenient to interpret the GW polarization tensor defined in eq.(\ref{Aij}) in terms of cosmological perturbations theory \cite{CPTLifshitz,Kodama:1985bj,Baumann:2009ds}. Since tensor perturbations are gauge invariant at first order, we will just consider scalar and vector perturbations. Using the scalar-vector-tensor (SVT) decomposition \cite{CPTLifshitz}, the scalar and vector perturbations of the flat FRW line element can be written as  
\be
\label{pmet}
\d s^2 = (1+2 \Phi) \d t^2 - 2 a(t) \Big(B_{,i}+S_{i}\Big) \d x^i \d t - a^2(t)\Big[(1-2\Psi) \delta_{ij} + 2 (E_{,ij}+ F_{(i, j)})\Big] \d x^i \d x^j\,,
\ee
where $S_{i}$ and $F_{i}$ are the vector perturbations, satisfying $S_{i,i}= F_{i,i} = 0$, and $\Phi,B,\Psi,E$ are the scalar perturbations.
Applying the SVT decomposition to an infinitesimal time and space translation, we obtain 
\bea
t &\to& t + T\ , \\
x^i &\to& x^i + \beta^i\, +\delta^{ij} \beta_{,j} \, .
\eea
where $\beta_{i}$ and  $\beta$ are respectively the vector and scalar part of the infinitesimal space translation, satisfying $\beta_{i,i}=0$, and $T$ is the infinitesimal time translation.

Under the above infinitesimal coordinate transformations the  metric tensor transformations implies the gauge transformations for the perturbations \cite{Baumann:2009ds}
\bea
S_i &\to& S_i + a \dot \beta_i\, , \\
F_i &\to& F_i - \beta_i\,, \\
\Phi &\to& \Phi - \dot T\,, \\
B &\to & B + a^{-1} T - a \dot \beta\,, \\
E &\to & E - \beta\,, \\
\Psi &\to& \Psi + H T \, .
\eea
while scalar fields transform as
\be
\delta\phi\rightarrow \delta\phi-\dot{\phi} \,T\,.
\ee
Note that the above gauge transformations  have a purely geometrical origin, i.e. they are independent of the specific theory, since they are  simply a direct consequence of the fact that the metric  and the energy-momentum tensor transform as tensors under a coordinate transformation. This allows to obtain a gauge invariant definition of GWs polarizations valid in any gravity theory, since the SVT decomposition of a tensor is also purely geometrical, and is not based on assuming any symmetry of the action defining the theory. In other words gauge transformations are just a manifestation of general covariance, which any theory should satisfy, and corresponds to the  Lagrangian not containing any free tonsorial index, i.e. being a coordinate invariant.
By comparing eq.(\ref{Aij}) with the spatial part of the perturbed metric we obtain:
\bea
h_S=1-2\Psi \quad&,&\quad h_L=1-2\Psi+E_{,zz} \,,\\
h_{V1}=-F_{x,z} \quad&,&\quad h_{V2}=-F_{y,z} \,,
\eea
from which we obtain the gauge transformations of the GWs polarizations
\bea
h_S&\to& h_S-2 H T \,, \\ 
h_L&\to& h_L-2 H T-\beta_{,zz} \,, \\
h_{V1} &\to& h_{V1} +\beta_{x,z} \,, \\
h_{V2} &\to& h_{V2} +\beta_{y,z}  \,.
\eea
We can now use the above gauge transformations to define gauge invariant GWs polarizations. It is convenient to fix the Einstein gauge (EG), defined by the condition 
\be
\nabla^{\mu}\Big(h_{\mu\nu}-\frac{1}{2}h g^0_{\mu\nu}\Big)=0
\ee
where $g^0_{\mu\nu}$ denotes the  background metric, 
$h=h^{ab} g^{0}_{ab}$ is the trace of $h_{ab}$,  and the covariant derivative is w.r.t. the background metric.
The above gauge fixing condition gives a set of four differential equations which can be solved to obtain the infinitesimal translations necessary to switch to the EG. Denoting the solutions of the gauge fixing equations as $\{T^{E},\beta^E,\beta_i^E\}$, we can define the gauge invariant GWs polarizations $\bar{h}_{ij}$
\bea
\bar{h}_S&=& h_S-2 H T^E \,, \\ 
\bar{h}_L&=& h_L-2 H T^E-\beta^E_{,zz} \,, \\
\bar{h}_{V1} &=& h_{V1} +\beta^E_{x,z} \,, \\
\bar{h}_{V2} &=& h_{V2} +\beta^E_{y,z}  \,,
\eea
which are gauge invariant by construction, and we will call EGWs. We are denoting with a bar the gauge invariant quantities EGWs defined above,  while tensor modes  are gauge invariant a first order in perturbations, i.e. $h_{\times}=\bar{h}_{\times},h_{+}=\bar{h}_{+}$. Some residual gauge freedom is present even after fixing the EG \cite{Flanagan:2005yc}, but we will not use it, since our main purpose is to define gauge invariant variables satisfying a wave equation, and  imposing the EG is enough to achieve this, as we will show in the next section.

\section{Gauge invariant gravitational waves equation}
The field equations of a generic JMC modified gravity theory defined by the action (\ref{JMCJ}) can be written in the Einstein frame as 
\be
G_{\E,\mu\nu}=T^{tot}_{\mu\nu}\label{EEEF}\,,
\ee
where $T^{tot}_{\mu\nu}$ is the sum of terms associated to the matter fields, the modified gravity fields, and their interaction with matter due to the non minimal coupling of matter with the Einstein frame metric.
The linearized perturbed Einstein equations w.r.t. to a curved background \cite{Ezquiaga:2018btd} in the Einstein gauge read
 \be
 \label{hCS}
\Box\psi_{\mu\nu}+2R^{_\text{B}}_{\mu\alpha\nu\beta}\psi^{\alpha\beta}=
  2\delta T_{\mu\nu}+2R^{_\text{B}~\alpha}_{~(\mu}\psi_{\nu)\alpha}-R^{_\text{B}}\psi_{\mu\nu}+g^0_{\mu\nu} R^{\alpha\beta}_{_\text{B}}\psi_{\alpha\beta}\,,
 \ee
 where $\psi_{\mu\nu}=\bar{h}_{\mu\nu}-\frac{1}{2}\bar{h} g^0_{\mu\nu}$.
 From the above equation we obtain
\be
\Box \bar{h}_{\mu\nu}=\delta T^{eff}_{\mu\nu}\label{hEG} \,,
\ee
where the d'Alambert operator is defined w.r.t. the background metric and $T^{eff}_{\mu\nu}$ is an effective energy-stress (ES) tensor given by the sum of the matter ES tensor, and other terms involving GWs, the Ricci tensor, and possible additional fields related to the gravity modification.
Note that the TT gauge cannot be imposed in a generic curved space \cite{Flanagan:2005yc}, and the scalar and vector modes cannot be gauged away in a generic modified gravity theory, since this is only possible in vacuo, but the extra fields associated to the gravity modification act as an effective source in the r.h.s. of eq.(\ref{hEG}), which in general is not zero even in absence of matter fields.
In GR, assuming a flat background, and in vacuo, the effective source term is zero, because the Ricci and Riemann  tensors are zero. This allows use the residual gauge freedom to set $\bar{h}=0$, and fix the TT gauge, in which only the two tensor modes survive. Note that eq.(\ref{hEG}) is quite general, since it is valid for any theory which admits an Einstein frame formulation, for a general background metric, and as mentioned earlier, the definitions of the gauge invariant variables is even more general, since it does not assume any form of the action, and the EG can be fixed even for a background different form the FRW solution.

Considering a FRW background, following a method similar to the one used in the previous section,  an effective equation for the gauge invariant GWs can be derived from eq.(\ref{hEG})
\be
\bar{h}_{ij}''+2 \Big(\frac{a'}{a}-\frac{c'_{ij}}{c_{ij}}\Big) \bar{h}'_{ij}+k^2 c_{ij}^2 \bar{h}_{ij}=0 \,,\label{heffbar}
\ee 
where
\be
c_{ij}^2(\eta,k)=\Big(1-\frac{\hat{g}_{ij}}{\hat{\bar{h}}'_{ij}}\Big)^{-1} \quad,\quad 
\hat{g}_{ij}=\frac{1}{a^2}\int a^4 \delta T^{eff}_{ij}\,d\eta\,, \label{cijbar}
\ee
Note that, contrary to the previous section derivation, which was based on assuming a non gauge invariant ansatz for the GWs propagation equation, the effective equation above is gauge invariant by construction and is quite general, since it can be obtained for any theory which admits an Einstein frame formulation. Formally it can also be applied to the solutions $\hat{g}_{\mu\nu}$ of theories whose field equations are not of the form given in eq.(\ref{EEEF}), by appropriately defining and effective ES tensor
\bea
F[\hat{g}_{\mu\nu}]&=&T_{\mu\nu}[\hat{g}_{\mu\nu},\hat{\phi}_i] \quad\,,\quad M_i[\hat{\phi}_i]=0\,,\label{Fg}\\
G_{\mu\nu}[\hat{g}_{\mu\nu}]&=&\hat{T}^{eff}_{\mu\nu}\,,\label{Gg} \\
\hat{T}^{eff}_{\mu\nu}(x^{\rho})&=&T_{\mu\nu}[\hat{g}_{\mu\nu},\hat{\phi}_i]-F[\hat{g}_{\mu\nu}]+G_{\mu\nu}[\hat{g}_{\mu\nu}]\,,
\eea
where $F$ and $M_i$ are the differential operator corresponding to the gravity and matter field equations, $\{\hat{g}_{\mu\nu},\hat{\phi}_i\}$ are solutions of eq.(\ref{Fg}), and  the components of $\hat{T}^{eff}_{\mu\nu}$ are functions of space and time, obtained by substituting into ${T}^{eff}_{\mu\nu}$ the solutions of the coupled  matter and gravity field equations (\ref{Fg}). 

Note that the above equations should be interpreted as a statement about the fact that any solution $\hat{g}_{\mu\nu}$ of eq.(\ref{Fg}) can also be obtained as a solution of eq.(\ref{Gg}), for an appropriate choice of $\hat{T}^{eff}_{\mu\nu}$ and boundary conditions, not as a statement about the full equivalence between  general relativity and the generic theory corresponding to eq.(\ref{Fg}), which may indeed involve different differential operators, and be fundamentally different. Since what we observe are the solutions of the field equations, not the equations themselves,
eq.(\ref{Gg}) is enough to obtain an effective description of the solutions of eq.(\ref{Fg}), according to the method given above.

\section{Generalization to higher order in perturbations}

Note that the gauge invariant quantities EGWs and eq.(\ref{hEG}) are defined at linear order in perturbations, and at higher order new gauge invariant variables ${\bar{h}}^{(i)}_{\mu\nu}$ can be defined \cite{Mollerach:2003nq,DeLuca:2019ufz,Chang:2020iji}. The expansion of the Einstein equations (\ref{EEEF}) will give new equations which can always be put in the canonical form
\be
\Box \bar{h}^{(i)}_{\mu\nu}=\delta T^{(i)eff}_{\mu\nu}\label{hEGn} \,,
\ee
by appropriately defining the effective perturbed ES tensor $\delta T^{(i)eff}_{\mu\nu}$, even if the d'Alambert operator does not appear explicitly, by adding it on both sides of the expanded equations.  We have fixed the gauge by imposing the condition 
\be
\nabla^{\mu}\Big(\bar{h}^{(i)}_{\mu\nu}-\frac{1}{2}\bar{h}^{(i)} g^0_{\mu\nu}\Big)=0\,,
\ee
where $g^0_{\mu\nu}$ denotes the  background metric, 
$\bar{h}^{(i)}=\bar{h}^{(i)}_{\mu\nu} g^{0,\mu\nu}$ is the trace of $\bar{h}^{(i)}_{\mu\nu}$,  and the covariant derivative is w.r.t. the background metric.
This effective equation allows to interpret GWs at any order in perturbations as the solutions of a wave equation with an appropriately defined source term.
Since the d'Alambert operator is defined w.r.t. to  the same background metric in all these equations, the equations can be summed to give
\be
\Box \bar{h}^{(N)}_{\mu\nu}=\delta T^{(N)eff}_{\mu\nu}\label{hEGN} \,.
\ee
where we have defined the summed GWs and ES tensor perturbations
\be
\bar{h}^{(N)}_{\mu\nu}=\sum^N_{i=1} \bar{h}^{(i)}_{\mu\nu} \quad,\quad T^{(N)eff}_{\mu\nu} =\sum^N_{i=1} \delta T^{(i)eff}_{\mu\nu}\label{hTN} \,.
\ee

Note that $\bar{h}^{(N)}_{\mu\nu}$ are the physically observable GWs, given by the sum of the contributions from different orders in perturbations.
From eq.(\ref{hEGN}), similarly to what shown in the previous section for linear perturbations, it is possible to define an effective speed, equation, and action for the summed GWs $\bar{h}^{(N)}_{\mu\nu}$. Note that this effective speed is not simply the sum of the effective speeds corresponding to perturbations equations at different orders, due to the coupling between perturbations at different orders. After solving the system of coupled differential equations for the perturbations equations at all relevant orders, the solutions $\hat{\bar{h}}^{(i)}_{\mu\nu}$  can be substituted to obtain $\hat{T}^{(N)eff}_{\mu\nu}$, from which  the  effective speed and equation can be obtained

\be
\bar{h}^{(N)''}_{ij}+2 \Big(\frac{a'}{a}-\frac{c'_{ij}}{c_{ij}}\Big) \bar{h}^{(N)'}_{ij}+k^2 c_{ij}^2 \bar{h}^{(N)}_{ij}=0 \,,\label{heffNbar}
\ee 

\be
c_{ij}^2(\eta,k)=\Bigg[1-\frac{\hat{g}_{ij}}{\hat{\bar{h}}_{ij}^{(N)'}}\Bigg]^{-1} \quad,\quad 
\hat{g}_{ij}=\frac{1}{a^2}\int a^4 \delta \hat{T}^{(N)eff}_{ij}\,d\eta\,. \label{cijN}
\ee
 The effective speed is encoding in a single quantity all the   interaction effects between different GWs polarizations, and between GWs and other fields, up to order $N$ in perturbations.

\section{Effective Lagrangian and metric}
The Lagrangian corresponding to eq.(\ref{heffNbar}) is
\be
\mathcal{L}^{eff}_h=\frac{a^2}{c_{ij}^2}\Bigg[ {\Big(\bar{h}^{(N)'}_{ij}\Big)}^2+ k^2 c_{ij}^2 {\Big(\bar{h}^{(N)}_{ij}\Big)}^2\Bigg],\label{L}
\ee
generalizing to scalar and vector modes the tensor perturbations effective action \cite{Romano:2022jeh}.
The effective method can also be applied in physical space, and it can be shown that \cite{Romano:2022jeh} the
effective Lagrangian  can be obtained from the GR  Lagrangian density

\be
\mathcal{L}_h^{GR}=a^2\Big[h'^2_{ij}-c^2 (\nabla h_{ij})^2\Big]=\sqrt{-g}(\partial_{\mu}h_{ij} \partial^{\mu}h_{ij})\,,
\ee
via the transformation
\be
a\rightarrow \alpha_{ij}=\frac{a}{\hat{c}_{ij}} \quad,\quad c\rightarrow \hat{c}_{ij} \quad,\quad h_{ij} \rightarrow \bar{h}^{(N)}_{ij} \,,\label{trans}
\ee
where we have denoted with $c$ the speed of light, to avoid ambiguity, and the space effective sound speed $\hat{c}_{ij}(\eta,x^i)$ \cite{Romano:2022jeh}, is defined in terms of the physical space effective ES tensor, not of its Fourier transform. Note that ${c}_{ij}(\eta,k)$ is not the Fourier transform of $\hat{c}_{ij}(\eta,x^i)$ \cite{Romano:2022jeh}.
The physical space effective action  is
\be
\mathcal{L}^{eff}_h=\sqrt{-g^{\rm{eff}}}(\partial_{\mu}h_{ij} \partial^{\mu}h_{ij}) \,,
\ee
where the effective metric is
\be
ds^2_{eff}=g^{\rm{eff}}_{\mu\nu}dx^{\mu}dx^{\nu}=a^2\Big[\hat{c}_{ij}d\eta^2-\frac{\delta_{mm}}{{\hat{c}_{ij}}}dx^mdx^m\Big] \label{geff} \,,
\ee
from which the EOM  are given in terms of the covariant d'Alembert operator 
\be
\square h_{ij}=\frac{1}{\sqrt{-g^{\rm{eff}}}}\partial_{\mu}\Big(\sqrt{-g^{\rm{eff}}}\partial^{\mu}h_{ij}\Big)=0 \,.
\ee
In this effective geometrical description, the effects of the interactions of GWs are encoded in the effective metric, and in the eikonal approximation,  the solutions of the EOM are geodesics in the effective curved space corresponding to the effective metric.

\section{Jordan frame effective action}
Cosmological perturbations w.r.t. to a flat FRW background are invariant under time dependent conformal transformations, since they correspond to a scale factor redefinition $a=\Omega\, \tilde{a}$, which has no effects on the perturbations 
\be
ds^2=a^2\Big[(\eta_{\mu\nu}+\delta g_{\mu\nu}) dx^{\mu}dx^{\nu}\Big]=\Omega^2(\eta) \tilde{a}^2\Big[(\eta_{\mu\nu}+\delta g_{\mu\nu}) dx^{\mu}dx^{\nu}\Big]\,.
\ee
The gauge invariant GWs do not depend on the coordinate choice by definition, and can in particular be evaluated in the unitary gauge, defined by $\delta\phi=0$, in which the conformal factor $\Omega$ becomes a function of time only. Using  the conformal and gauge invariance of the GWs $\bar{h}_{ij}$, we can obtain the Jordan frame effective action in the unitary gauge   
\be
\mathcal{L}^{eff}_{\bar{h}}= \frac{\Omega^2(\eta)\tilde{a}^2}{\hat{c}_{ij}^2}\Big[ \bar{h}'^2_{ij}-  \hat{c}_{ij}^2 (\nabla \bar{h}_{ij})^2\Big]= \hat{M}^2_{ij}\tilde{a}^2\Big[ \bar{h}'^2_{ij}-  \hat{c}_{ij}^2 (\nabla \bar{h}_{ij})^2\Big]\label{Lxbar}\,.
\ee
where $\hat{M}_{ij}=\Omega/\hat{c}_{ij}$ plays the role of space and polarization dependent effective Planck mass. 

\section{Polarization and frequency dependency of the luminosity distance}
After defining the effective scale factor
\be
\alpha_{ij}=\frac{a}{c_{ij}} \,,
\ee
equation (\ref{heffNbar}) can be re-written as
\be
\bar{h}^{(N)''}_{ij}+2 \frac{\alpha_{ij}'}{\alpha_{ij}} \bar{h}^{(N)'}_{ij}+k^2 c_{ij}^2 \bar{h}^{(N)}_{ij}=0 \label{heffNa}\,,
\ee 
from which we  obtain 
\be
\chi_{ij}''+\Big(c_{ij} k^2 -\frac{\alpha_{ij}''}{\alpha_{ij}}\Big)\chi_{ij}=0 \,,
\ee
where we have defined $\bar{h}^{(N)}_{ij}=\chi_{ij}/\alpha_{ij}$.
In the sub-horizon limit $\alpha_{ij}''/\alpha_{ij}$ is negligible, and the amplitude of $\bar{h}^{(N)}_{ij}$ is proportional to $1/\alpha_{ij}$, giving \cite{Romano:2022jeh}
\be
\frac{d_{ij}^{GW}}{d_{L}^{EM}}(z)=\frac{a(z)}{\alpha_{ij}(z)}\frac{\alpha_{ij}(0)}{a(0)}=\frac{c_{ij}(z,k)}{c_{ij}(0,k)}\,,\label{dgw}
\ee
where we have used $d_{ij}^{GW}=r\,\alpha_{ij}(0)/\alpha_{ij}(z)$, $d_L^{EM}=r\, a(0)/a (z)$, assuming $(1+z)=a(0)/a(z)$, i.e. that  matter is minimally coupled to the Einstein frame metric.

If matter is minimally coupled to the Jordan frame metric, eq.(\ref{heffNa}) is still valid, due to the conformal invariance of GWs, but the relation between the scale factor and $\alpha$ is modified 
\be
\alpha_{ij}=\frac{a}{c_{ij}}=\frac{\Omega \tilde{a}}{c_{ij}} \,,
\ee
implying that
\be
\frac{d_{ij}^{GW}}{d_{L}^{EM}}(z)=\frac{\tilde{a}(z)}{\alpha_{ij}(z)}\frac{\alpha_{ij}(0)}{\tilde{a}(0)}=\frac{c_{ij}(z,k)}{c_{ij}(0,k)} \frac{\Omega(0)}{\Omega(z)}=\frac{M_{ij}(0,k)}{M_{ij}(z,k)}\,,\label{dgw}
\ee
where we have used $d_{ij}^{GW}=r\,\alpha_{ij}(0)/\alpha_{ij}(z)$, $d_L^{EM}=r\, \tilde{a}(0)/\tilde{a}(z)$, assuming $(1+z)=\tilde{a}(0)/\tilde{a}(z)$, i.e. that  matter is minimally coupled to the Jordan frame metric.
The quantity $M_{ij}=\Omega/c_{ij}$ plays the role of polarization and momentum dependent effective Planck mass. 

Note that independently of the type of matter-gravity coupling, the GWs luminosity distance is predicted to be frequency and polarization dependent.

\section{Observational implications}
The polarization and frequency dependency of the effective speed implies that different polarizations of GWs emitted by the same source at different frequencies can spend different times to reach the observer.
This effect could be detected  by comparing the time delay between  the detection of different GWs polarizations with different observatories, operating at different frequencies, and can be observed even in absence of an electromagnetic counterpart.

Another observable effect is the polarization and frequency dependency of the GWs luminosity distance. 
For theories with matter minimally coupled to the Jordan frame metric the GW luminosity distance  is related to the effective Planck mass ratio
\be
d_{ij}^{GW}(z)=\frac{M_{ij}(0,k)}{M_{ij}(z,k)}d_L^{EM}(z)\,,\label{dgwOM}
\ee
while for  theories minimally coupled to the Einstein frame is related to the effective speed ratio
\be
d_{ij}^{GW}(z)=\frac{c_{ij}(0,k)}{c_{ij}(z,k)}d_L^{EM}(z)\,.\label{dgwOC}
\ee
Time delay observations allow to constrain at different frequencies the ratio between the speed of different GWs polarizations 
\be
r^c_{ijpq}(k,z)=\frac{c_{ij}(k,z)}{c_{pq}(k,z)}\,,
\ee while the GWs waveform, which are inversely proportional to the GW luminosity distance, allow to constrain the corresponding distance ratio 
\be
r^d_{ijpq}(k,z)=\frac{d_{ij}^{GW}(z)}{d_{pq}^{GW}(z)}.
\ee
For multimessenger events additional constraints can be set on $c_{ij}(k,z)/c$ and $d_{ij}^{GW}(z)/d^{EM}_L(z)$.

For GWs propagating in vacuum according to GR, only  tensor modes are expected, and there should not be any redshift, frequency or polarization dependency, i.e. $c_{\times}(k,z)=c_{+}(k,z)=c$ and $d_{\times}^{GW}(z)=d_{+}^{GW}(z)=d^{EM}_L(z)$ at any frequency. If the effective ES tensor defined in eq.(\ref{hTN}) is not negligible along the GW propagation, either because of GWs interaction with matter fields, or due to the effects of gravity modification, other polarizations modes could be detected and the corresponding speed and luminosity distance  could be constrained observationally using time delay and GWs waveform observations. These observations can be used to probe the dark Universe by its  interaction with GWs, modeled by the effective GWs speed.

\section{Conclusions}
We have derived a gauge invariant effective equation and action for GWs, encoding the effects of interaction at different orders in perturbations theory in a polarization, frequency and time dependent effective speed. The invariance of perturbations under time dependent conformal transformations and  the gauge invariance of the GWs allow to   obtain the unitary gauge effective action in any conformally related frame. 
The propagation time and luminosity distance of different GWs polarizations, emitted at different frequencies and redshifts by  dark or bright sirens, allow to probe the interaction with other fields, and in particular with the dark Universe.

In the future it will important compute the effective GWs speed for different dark matter and dark energy theories, to constrain these theories with luminosity distance and time delay observations.

\begin{acknowledgments}
I thank Antonaldo Diaferio, Francesco Pace e Piero Rettegno and for interesting discussions regarding GWs polarizations. I thank the YITP, and Osaka University Theoretical Astrophysics group for the kind hospitality during the preparation of part of this paper.
\end{acknowledgments}

\bibliographystyle{h-physrev4.bst}
\bibliography{Bibliography.bib}

\begin{thebibliography}{10}

\bibitem{LIGOScientific:2016aoc}
LIGO Scientific, Virgo, B.~P. Abbott {\em et~al.},
\newblock Phys. Rev. Lett. {\bf 116}, 061102 (2016), arXiv:1602.03837.

\bibitem{Romano:2022jeh}
A.~E. Romano,
\newblock Phys. Lett. B  (2024), arXiv:2211.05760.

\bibitem{Romano:2023ozy}
A.~E. Romano and M.~Sakellariadou,
\newblock Phys. Rev. Lett. {\bf 130}, 231401 (2023), arXiv:2302.05413.

\bibitem{Romano:2018frb}
A.~E. Romano and S.~A. Vallejo~Pena,
\newblock Phys. Lett. B {\bf 784}, 367 (2018), arXiv:1806.01941.

\bibitem{Romano:2023uwf}
A.~E. Romano,
\newblock (2023), arXiv:2301.05679.

\bibitem{Romano:2023lxf}
A.~E. Romano and S.~A. Vallejo-Pe\~na,
\newblock (2023), arXiv:2301.11304.

\bibitem{Ezquiaga:2018btd}
J.~M. Ezquiaga and M.~Zumalac\'arregui,
\newblock Front. Astron. Space Sci. {\bf 5}, 44 (2018), arXiv:1807.09241.

\bibitem{Bonvin:2022mkw}
C.~Bonvin {\em et~al.},
\newblock Mon. Not. Roy. Astron. Soc. {\bf 525}, 476 (2023), arXiv:2211.14183.

\bibitem{Nishizawa:2017nef}
A.~Nishizawa,
\newblock Phys. Rev. D {\bf 97}, 104037 (2018), arXiv:1710.04825.

\bibitem{Creminelli:2014wna}
P.~Creminelli, J.~Gleyzes, J.~Nore\~na, and F.~Vernizzi,
\newblock Phys. Rev. Lett. {\bf 113}, 231301 (2014), arXiv:1407.8439.

\bibitem{Gubitosi:2012hu}
G.~Gubitosi, F.~Piazza, and F.~Vernizzi,
\newblock JCAP {\bf 02}, 032 (2013), arXiv:1210.0201.

\bibitem{CPTLifshitz}
E.~Lifshitz,
\newblock Zh. Eksp. Teor. Phys. {\bf 16} {\bf 16}, 587 (1946).

\bibitem{Kodama:1985bj}
H.~Kodama and M.~Sasaki,
\newblock Prog. Theor. Phys. Suppl. {\bf 78}, 1 (1984).

\bibitem{Baumann:2009ds}
D.~Baumann,
\newblock {Inflation},
\newblock in {\em {Theoretical Advanced Study Institute in Elementary Particle Physics}: {Physics of the Large and the Small}}, pp. 523--686, 2011, arXiv:0907.5424.

\bibitem{Flanagan:2005yc}
E.~E. Flanagan and S.~A. Hughes,
\newblock New J. Phys. {\bf 7}, 204 (2005), arXiv:gr-qc/0501041.

\bibitem{Mollerach:2003nq}
S.~Mollerach, D.~Harari, and S.~Matarrese,
\newblock Phys. Rev. D {\bf 69}, 063002 (2004), arXiv:astro-ph/0310711.

\bibitem{DeLuca:2019ufz}
V.~De~Luca, G.~Franciolini, A.~Kehagias, and A.~Riotto,
\newblock JCAP {\bf 03}, 014 (2020), arXiv:1911.09689.

\bibitem{Chang:2020iji}
Z.~Chang, S.~Wang, and Q.-H. Zhu,
\newblock (2020), arXiv:2009.11994.

\end{thebibliography}

\end{document}